\documentclass[journal]{IEEEtran}
\usepackage{graphicx}
\usepackage{multicol}
\usepackage[inline,shortlabels]{enumitem}
\usepackage{cite}
\usepackage{subfigure}
\usepackage{amsmath, amssymb, amsthm, amsfonts, latexsym, bm, graphicx, rawfonts, subfigure, multirow, rotating, float, array, color}
\usepackage[table]{xcolor}
\usepackage{booktabs}
\usepackage{MnSymbol}

\def\a{{\boldsymbol a}} 
\def\b{{\boldsymbol b}} 
\def\c{{\boldsymbol c}} 
 
\def\e{{\boldsymbol e}}

\def\j{{\boldsymbol j}} \def\J{{\boldsymbol J}}

\def\p{{\boldsymbol p}} 
\def\q{{\boldsymbol q}}

\def\t{{\boldsymbol t}}

\def\x{{\boldsymbol x}} \def\X{{\boldsymbol X}}
\def\y{{\boldsymbol y}} \def\Y{{\boldsymbol Y}}

\def\ba{{\boldsymbol{\alpha}}}
\def\bl{{\boldsymbol{\lambda}}}

\DeclareMathOperator*{\argmax}{argmax}

\newcommand{\subparagraph}{}
\usepackage{titlesec}
\titlespacing*{\subsection}{0pt}{0.5\baselineskip}{0.2\baselineskip}
\titlespacing*{\section}{0pt}{0.5\baselineskip}{0.2\baselineskip}

\begin{document}
%
\title{Information-Theoretic Interactive Sensing and Inference for Autonomous Systems}
%
%
%

\author{Christopher~Robbiano,~\IEEEmembership{Member,~IEEE,}
        Mahmood~R.~Azimi-Sadjadi,~\IEEEmembership{Life~Member,~IEEE,}
        ~and~Edwin~K.~P.~Chong,~\IEEEmembership{Fellow,~IEEE}
\thanks{C. Robbiano, M.R. Azimi-Sadjadi, E.K.P. Chong, are with the Electrical and Computer Engineering department at Colorado State University, Fort Collins, CO 80524 USA (email: \{chris.robbiano, edwin.chong, azimi\}@colostate.edu).}
\thanks{This work was supported by the Office of Naval Research (ONR) under contract N00014-18-1-2805.}%
\thanks{Manuscript received March ??, 2020; revised Month ??, 2020.}}

\markboth{}%
{Robbiano \MakeLowercase{\textit{et al.}}: Information-Theoretic Approach to Navigation for Efficient Detection and Classification of Underwater Objects}


\maketitle

\begin{abstract}
This paper addresses an autonomous exploration problem in which a mobile sensor, placed in a previously unseen search area, utilizes an information-theoretic navigation cost function to dynamically select the next sensing action, i.e., location from which to take a measurement, to efficiently detect and classify objects of interest within the area.  The information-theoretic cost function proposed in this paper consist of two \textit{information gain} terms, one for detection and localization of objects and the other for sequential classification of the detected objects. We present a novel closed-form representation for the cost function, derived from the definition of mutual information. We evaluate three different policies for choosing the next sensing action: lawn mower, greedy, and non-greedy. For these three policies, we compare the results from our information-theoretic cost functions to the results of other information-theoretic inspired cost functions. Our simulation results show that search efficiency is greater using the proposed cost functions compared to those of the other methods, and that the greedy and non-greedy policies outperform the lawn mower policy.
\end{abstract}


%
\IEEEpeerreviewmaketitle

\section{Introduction}
%
%
%
%
\IEEEPARstart{I}{n} this paper, we consider the problem of autonomous exploration for the purpose of interactive sensing and inference in previously unseen search areas.  At each time step, the autonomous platform performs a sensing action in the form of selecting and moving to the next position to collect a measurement that is used to update the detection, localization, and classification estimates. In this exploration problem, often referred to as the \textit{active perception} problem \cite{bajcsy1988active}, no pre-planned platform path is assumed as there is no \textit{a priori} information about objects in the search area, and all initial sensing actions are regarded as providing the same amount of information.  Additionally, the motion of the platform is restricted by some dynamical model, hence precluding arbitrary sequential sensing locations.

In active perception problems, the efficiency with which the search area is surveyed is typically the most important criterion as it directly relates to operational costs per sensing action, a need to minimize surveillance time during an information gathering sortie, as well as other time- and cost-sensitive objectives.  That is, one is concerned with achieving high efficiency through minimizing the number of sensing actions, while maximizing the detection and classification performance. 

To achieve such goals, information-theoretic measures have typically been used \cite{whaite1997autonomous, hollinger2013active, bourgault2002information, bai2018inference, julian2014mutual} for choosing optimal sensing actions in autonomous navigation and exploration problems.  In the case of parameter estimation using measurements that are corrupted by Gaussian noise, maximizing the Shannon entropy (amount of information) of the error distribution is equivalent to minimizing the determinant of the parameter estimate covariances \cite{whaite1997autonomous}.  This provides a rule for selecting the sensing action that maximizes the predicted variance of the measurement produced after a sensing action is performed.

In \cite{hollinger2013active}, a subclass of the active perception problem is addressed, where an autonomous underwater vehicle (AUV) is used to inspect the hull of a large ship and estimate its surface shape.  Gaussian process function approximation is exploited to approximate a mutual information-based cost function.  In this particular multi-hypothesis testing problem, \textit{a priori} information is available that allows the entire set of sensing actions, and their outcomes, to be observed prior to visiting all sensing locations.  

Information-theoretic cost functions, specifically utilizing \textit{information gain}, have been previously developed \cite{bourgault2002information,bai2018inference,julian2014mutual}, and used successfully, in the context of navigation using information from the occupancy grid estimation process \cite{elfes1989using}, \cite{thrun2003learning}.  

In \cite{bourgault2002information}, a measurement is estimated for a given sensing action using an extended Kalman filter (EKF), and then the mutual information is directly calculated following an update to the occupancy grid using the estimated measurement. In addition to the occupancy grid based information gain cost, \cite{bourgault2002information} suggests formulating an additional information-theoretic cost function from the outputs of the simultaneous localization and mapping (SLAM) problem using an EKF to estimate the positions of the AUV and objects in the search area.  Specifically, the cost function they choose is related to the determinant of the error covariance matrices for the AUV and objects, similar to \cite{whaite1997autonomous}.  A convex combination of the two normalized cost functions is used in the sensing action selection, providing the ability to trade-off performance in localization (through SLAM) and detection (through occupancy grids) of objects. 

In \cite{bai2018inference}, the mutual information is directly calculated after each sensing action and subsequent measurement is taken \textit{a priori}, and used to train the Gaussian process regression network for estimating the mutual information for future sensing actions.  Bayesian optimization is then used in conjunction with the Gaussian process upper confidence bound to estimate the information gain for each point in an occupancy grid. 

The formulation for explicitly calculating the predicted mutual information in \cite{julian2014mutual} is developed using an occupancy grid framework under the assumption of statistical independence of measurements. Measurements are also assumed to be conditionally independent of the occupancy state of obscured grid cells, i.e. grid cells behind occupied cells in the perceptual range of the sensor, given an occupied grid cell.

In this paper, we propose a new approach to the problem of active perception using two information-theoretic cost functions based on information gain.  The first cost function is associated with object detection and localization, and measures the mutual information between the occupancy state variable for a single grid cell and a binary measurement random variable. Solving for its closed-form representation relies on the measurement model and posterior occupancy distributions produced through the occupancy grid estimation process presented in \cite{robbiano2020bayesian}.  The second cost function associated with the classification of detected objects measures the mutual information between a class state variable for a single grid cell and random variable that is the parameter to the class state variable distribution.  In this formulation, we choose to model the class state variable as a Categorical random variable, and its distribution parameter as a Dirichlet random variable \cite{murphy2012machine}.  The motivation  for choosing this modeling scheme comes from the need to perform sequential updating of the class state variable distribution, akin to occupancy grid estimation process.  This sequential updating process has a closed-form due to the conjugacy between the Dirichlet and Categorical distributions, and allows for fast tracking of the class state distribution as new measurement are drawn.  A \textit{one-step} classification process is used in our formulation, producing class labels used to sequentially update the class state variable distribution.  Similar to \cite{bourgault2002information}, a convex combination weighting of two normalized cost functions for sensing action selection is also utilized here. 

A series of experiments are conducted to illustrate the utility of the proposed sequential state updating in conjunction with the proposed cost functions.  Three sensing action selection policies are compared---lawn mower, greedy, and non-greedy.  The lawn mower policy does not use a cost function in the choice of the next sensing action.  For the two policies that use a cost function for selection the next sensing action, greedy and non-greedy, two different methods for estimating the information gain are used: 
\begin{enumerate*}[(a)]
    \item the convex combination of our two proposed cost functions, and
    \item a Gaussian process function approximation method.
\end{enumerate*}
The performance of all three policies is evaluated for their ability to explore the interrogation area, and detect and classify targets, while performing only a limited number of actions.  The results show that policies using the convex combination of our cost functions outperform all other policies without the need for any training data.

This paper is organized as follows. A review of the occupancy grid estimation process is presented in Section \ref{sec:occ_grids}. The sequential classification process, including a description of the one-step classifier, is presented in Section \ref{sec:seq_class}.  The  derivations of the information-theoretic cost functions for detection and classification are presented in Section \ref{sec:info_metrics}.  The different navigation policies are described in Section \ref{sec:nav_policies}.  Simulation results on synthetic sonar data, and a comparison with other cost functions are presented in Section \ref{sec:simulations}.  Finally, concluding remarks are made in Section \ref{sec:conc}.

\section{System Overview} \label{sec:system}
The specific active perception problem considered here involves undersea mine hunting in littoral zones using an AUV equipped with a side-looking sonar system, though the proposed formulations are not restricted only to this sensor configuration. The AUV explores previously unseen areas and simultaneously performs detection and classification of undersea objects.  
We divide this active perception problem into four segments: 
\begin{enumerate}[(1)]
    \item Generating a map of the scene through an occupancy grid estimation process, which produces a set of marginal posterior probabilities that any one point in an area is occupied \cite{robbiano2020bayesian}.
    \item Classifying  occupied regions through a sequential classification process, which produces a set of marginal posterior probabilities of class membership for each occupied region.
    \item Computing the 
        \begin{enumerate*}
            \item mutual information between the occupancy state of a grid cell and a random variable modeling a measurement on that grid cell and
            \item mutual information between the class state of a grid cell and a random variable modeling the class distribution parameter.
        \end{enumerate*}
        \item Exploiting the information gain to select the sensing action that produces the best opportunity to detect, localize, and classify an object.
\end{enumerate}

\begin{figure}
    \centering
    \includegraphics[width=.48\textwidth]{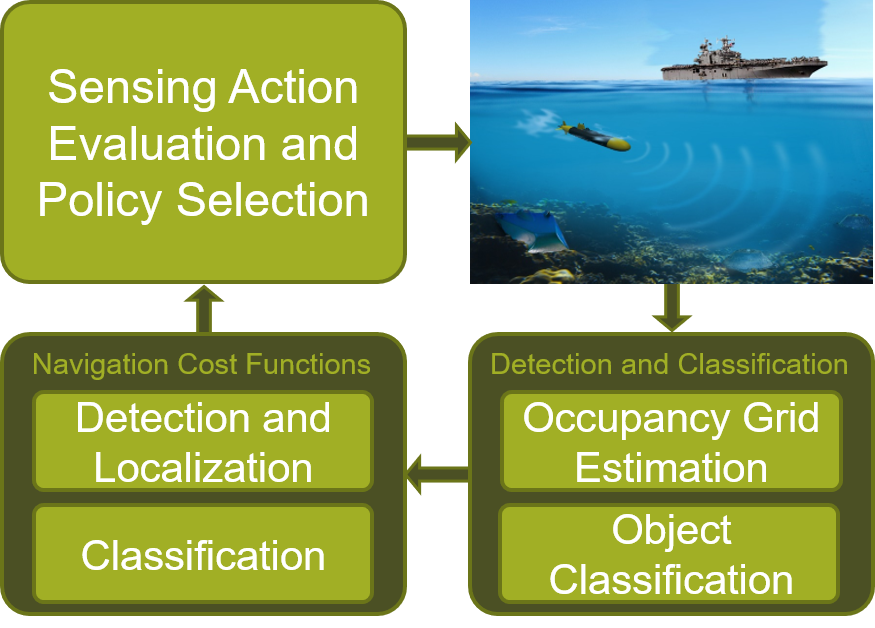}
    \caption{The proposed active perception problem: An AUV takes measurements to perform occupancy grid estimation and sequential classification.  The outputs from each of these processes are then used in evaluating the navigation cost function for sensing action and policy selection.}
    \label{fig:active_perception_problem}
\end{figure}

An illustration of the proposed active perception problem is given in Figure \ref{fig:active_perception_problem}.  
The search area is discritized into grid cells where the knowledge about object locations in the environment is captured through an occupancy grid estimation process \cite{robbiano2020bayesian}. The knowledge about the class of each detected and localized object is then provided in the form of a Dirichlet-Categorical model (DCM) \cite{murphy2012machine}.  The occupancy grid estimation process and the DCM produce a set of distributions over occupancy state and class state of grid cells, respectively, from which the uncertainty in the distributions can be measured through Shannon entropy \cite{cover2012elements}. The state distribution of each grid cell is updated using the measurement collected after each sensing action, and the reduction in the uncertainty of each distribution after each sensing action can be measured through \textit{information gain} or \textit{mutual information} \cite{cover2012elements}.  The information gain is expressed as the difference between the entropy of a prior distribution for a state variable and the entropy of the posterior distribution for that state variable after a measurement has been observed.  It is natural to seek sensing actions that will produce a measurement maximizing the reduction in uncertainty in both occupancy state and class state for all grid cells, and thus we choose to utilize information gain as our information-theoretic cost function for choosing sensing actions. These components of the proposed system are described in the next sections. 

In the remainder of this paper, we shall use lowercase italic $x$ for scalars, lowercase bold italic symbols $\x$, and uppercase bold italic symbols $\X$, for vectors and matrices, respectively.

\section{Occupancy Grid Estimation}
\label{sec:occ_grids}
Occupancy grid estimation is a popular process for generating an occupancy map of an area given a set of measurements taken from that area \cite{elfes1989using, thrun2003learning, robbiano2020bayesian}.  \sloppy The map is partitioned into a set of $B$ disjoint \textit{grid cells} $\{g_i\}_{i=1}^B$, all with the same shape and size. To each grid cell a binary occupancy state indicator variable $b_i\in\{0,1\}$ is attached  with $b_i=1$ indicating that a grid cell $g_i$ is \textit{occupied} (e.g., by a scatterer of radiation), and $b_i=0$ indicating that $g_i$ is \textit{empty}.  We call the set $\b = \{b_i\}_{i=1}^B$ the set of \textit{cellular occupancies}, commonly referred to as a \textit{map}.  The map $\b $ can be any one of $2^{B}$ possible unique maps from the set of all possible maps $\mathbb{B}$. 

\begin{figure}[h]
    \centering
    \includegraphics[width=.48\textwidth]{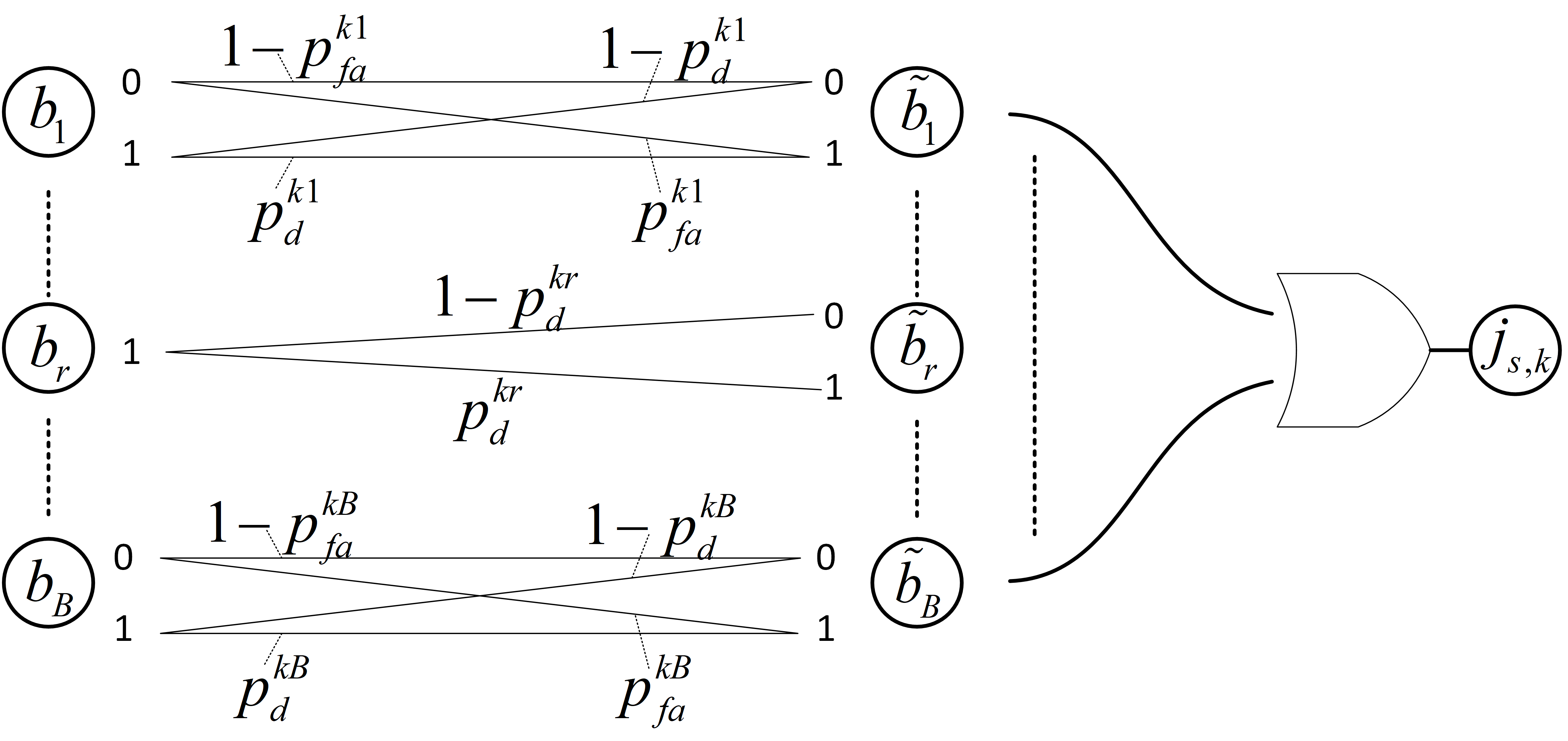}
    \caption{Modeling of interaction between occupancy states and a single measurement $j_{s,k}$, conditioned on $b_r=1$.}
    \label{fig:sensor_model}
\end{figure}

Now, given the measurement matrix $\J_S = \begin{bmatrix} \j_1, \hdots, \j_s, \hdots, \j_S \end{bmatrix}$, consisting of a collection of measurement vectors $\j_s = \begin{bmatrix} j_{s,1},\hdots,j_{s,K} \end{bmatrix} \in \mathcal{J}^K=\{0,1\}^K$ for $s\in\{1,\hdots, S\}$ with $K$ elements that are the thresholded range measurements taken at time $s$, the estimation problem produces the set of marginal posterior probabilities, or occupancy grids (OGs), arranged as a vector
\begin{align} \label{eq:og_posterior_prob}
    \p = \{p_{b|\J}(b_r = 1 | \J_S)\}_{r=1}^B.
\end{align}
Using the following occupancy grid formulation presented in \cite{robbiano2020bayesian}, these marginal posterior probabilities at time step $S$ can be expressed as
\begin{align} \label{eq:gridcellmarginal} \notag
    p_{b|\J}&(b_r = 1 | \J_S) \propto \sum_{\b\in \mathbb{B}(r,1)} p_{\j|\b}(\j_S| \b) p_{\b|\J}( \b|\J_{S-1}) \\ \notag
    &= \sum_{\b\in \mathbb{B}(r,1)}  \prod_k \prod_{i}\Big[ \big(p^{00}_{ki}(1-b_{i}) + p^{01}_{ki}b_{i}\big)(1-j_{S,k}) \\ 
    & \hspace{.3in} + \big(1-(p^{00}_{ki}(1-b_{i} + p^{01}_{ki}b_{i})\big)j_{S,k} \Big] \\ \notag
    &\hspace{.3in} \times p_{\b|\J}(\b | \J_{S-1}),
\end{align}
where $\mathbb{B}(r,1)$ is the set of all maps with the $r$th occupancy state pinned to occupied.  Given the map $\b$, for any arbitrary time $s\in [1, S]$, the \textit{sensor model} $p(\j_s|\b)$ can be written as
\begin{align} \label{eq:sensor_model}
    p_{\j|\b}(\j_s|\b) &= \prod_k p_{j|\b}(j_{s,k}|\b).
\end{align} 
To express the terms under the product in (\ref{eq:sensor_model}) the BAC model was adopted in \cite{robbiano2020bayesian}. Figure \ref{fig:sensor_model} illustrates the interaction between all of the grid cell occupancy states and a single measurement where each $j_{s,k}$ is a Boolean function of virtual occupancies $\tilde{b}_i$ (outputs of the BACs); specifically $j_{s,k} = \sum_{i=1}^{B} \tilde{b}_i$. Then, we can write, 
\begin{align} \label{eq:measurement_model} \notag
    p_{j|\b}(j_{s,k} = 0|\b) &= \prod_{i} p_{\tilde{b}|b}( \tilde{b}_{i} = 0 | b_{i}) \\
    &=  \prod_{i} p^{00}_{ki}(1-b_{i}) + p^{01}_{ki}b_{i},
\end{align} 
The quantity $p^{00}_{ki}$ is the probability that the occupancy state of grid cell $g_i$ is transmitted through the BAC and correctly received as measurement $j_{s,k}=0$ when $b_i=0$ (probability of true non-detection), and $p^{01}_{ki}$ is the probability that the occupancy state of grid cell $g_i$ is transmitted through the BAC and incorrectly received as measurement $j_{s,k}=0$ when $b_i=1$ (probability of missed detection).  As $j_{s,k}$ is a binary random variable, $p_{j|\b}(j_{s,k} = 1|\b) = 1-p_{j|\b}(j_{s,k} = 0|\b)$. The last term in (\ref{eq:gridcellmarginal}) $p_{\b|\J}(\b | \J_{S-1})$ is the posterior probability of the map $\b$ at the previous time step $S-1$ calculated as
\begin{align} \label{eq:og_joint_computable} \notag
    p_{\b|\J}(\b | \J_{S-1}) &\propto \prod_k \prod_{i}\Big[ \big(p^{00}_{ki}(1-b_{i}) + p^{01}_{ki}b_{i}\big)(1-j_{S-1,k}) \\ \notag
    & \hspace{.3in} + \big(1-(p^{00}_{ki}(1-b_{i}) + p^{01}_{ki}b_{i})\big)j_{S-1,k} \Big] \\
    &\hspace{.3in} \times p_{\b|\J}(\b | \J_{S-2}).
\end{align}

One method for choosing the BAC transition probabilities $p_{ki}^{00}$ and $p_{ki}^{01}$ is to allow $p_{ki}^{00} = (1-p_{\text{d}})/(1+\text{dist}(b_i,j_{s,k}))^{\alpha}$ and $p_{ki}^{01} = (1-p_{\text{fa}})/(1+\text{dist}(b_i,j_{s,k}))^{\alpha}$, where $p_{\text{d}}$ and $p_{\text{fa}}$ are the probability of detection and false alarm, respectively, of the physical sonar system, $\text{dist}(b_i, j_{s,k})$ represents the Euclidean distance between the location of grid cell $g_i$ and that at which sample $j_{s,k}$ was taken, and $\alpha \geq 1$ \cite{robbiano2020bayesian}. This particular modeling is used to emulate degraded detection performance due to attenuation in the sonar return signal strength as a function of distance.

This formulation of occupancy grid estimation is used over other estimation techniques as it is able to account for the correlation between occupancy states of neighboring grid cells, and was developed with the measurement type (binary detection statistics) that are used in the implementation of our system.

\section{Sequential Classification Using Dirichlet-Categorical Models}
\label{sec:seq_class}
In this section, we present a new sequential classification method that allows tracking of the class state for each grid cell, formulated with the need of an information-theoretic classification cost function in mind.  We desire a method that produces a set of distributions as its output, and the sequential updating of these distributions to be performed quickly, without the need to iterate an algorithm to convergence.  As such, we chose the Dirichlet-Categorical model (DCM) \cite{tu2014dirichlet, murphy2012machine} for representing the class state variable and its associated distribution random variable.  

The idea behind this sequential updating process is to take a measurement from the sensor at time $s$ and use it to update the class membership probabilities for the grid cell.  The measurement is first converted into a crude estimate of the class label $l_s$, and that label is merged with all previous labels to update the posterior predictive class distribution for the grid cell.  

Let $c$ be the class state variable for grid cell $g_i$.  At each time step $s$, a \textit{one-step} classifier is employed to assign a class label $l_s \in [1,L]$ to the most recent measurement from grid cell $g_i$. The one-step classifier in this system can be any commonly used classifier such as support vector machine (SVM), relevance vector machine (RVM) \cite{bishop2006pattern} or a deep neural network (DNN) \cite{goodfellow2016deep}.   The collection of sequential class labels $l_s$ for grid cell $g_i$ up to the current sensing time $S$, are formed into a set $\mathcal{L}=\begin{bmatrix} l_1, \hdots, l_S \end{bmatrix}$. Now, the goal here is to generate the posterior predictive distribution of the class state variable $c$, $p_{c|\mathcal{L}}(c|\mathcal{L})$, given all the past and present labels in $\mathcal{L}$. 

To begin, we model $c$ as a Categorical random variable taking on $L$ possible, non-orderable, values. A random variable $c$ is Categorically distributed if $p_{c|\bl}(c=l|\bl)=\lambda_l = P(c=l)$ for $l=1,\hdots, L$, where $\bl = \begin{bmatrix} \lambda_1, \hdots, \lambda_L \end{bmatrix}$ and $\sum_{l=1}^L \lambda_l = 1$, and can be expressed as $c|\bl \sim \text{Cat}(\bl)$\cite{murphy2012machine}. The probability mass function of the Categorical distribution can be written as 
\begin{align} \label{eq:catpdf}
    p_{c|\bl}(c|\bl) = \prod_{l=1}^L \lambda_l^{\delta_{cl}}, && \delta_{cl} = \begin{cases} 1 & c=l \\ 0 & \text{otherwise}  \end{cases}.
\end{align}
The Categorical distribution parameter $\bl$ is modeled as a Dirichlet distributed random variable with distribution parameter $\ba$, $\bl\sim\text{Dir}(\ba)$. The probability density function of $\bl$ is defined as
\begin{align} \label{eq:dirpdf}
    p(\bl) &=  \frac{1}{B(\ba)}\prod_{l=1}^L \lambda_l^{\alpha_l - 1},
\end{align}
where $B(\ba)=\frac{\prod_{l=1}^L \Gamma(\alpha_l)}{\Gamma(\alpha_0)}$ is the multivariate beta function and $\alpha_0 = \sum_{l=1}^L \alpha_l$ \cite{murphy2012machine}.  The parameter vector $\ba = \begin{bmatrix} \alpha_1, \hdots, \alpha_L \end{bmatrix}$ is non-random, with $\alpha_l>0~\forall~l$.

The Dirichlet distribution is the conjugate prior for the Categorical distribution, and thus the posterior distribution of $\bl|c$ is $\text{Dir}(\ba^{\circ})$ where $\ba^{\circ} = \begin{bmatrix} \alpha_1^{\circ}, \hdots, \alpha_L^{\circ} \end{bmatrix}$ and $\alpha_l^{\circ}=\alpha_l - 1 + \delta_{cl}$ \cite{murphy2012machine,tu2014dirichlet}. That is, we can write $\bl|c=\bl^{\circ}\sim\text{Dir}(\ba^{\circ})$. This shows that after getting a new class label $l_s$, the updated estimate of the distribution parameter $\bl$ is now Dirichlet distributed with parameter $\ba^{\circ}$.  We call this updated distribution parameter $\bl^{\circ}\sim\text{Dir}(\ba^{\circ})$.

The DCM provides an efficient closed-form equation for calculating the posterior predictive distribution \cite{tu2014dirichlet} of the class state variable $c$ given the label data in $\mathcal{L}$ using
\begin{align*}
    p_{c|\mathcal{L}}(c|\mathcal{L}) &= \int p_{c|\bl}(c|\bl) p_{\bl | \mathcal{L}}(\bl|\mathcal{L}) d\bl \\
    &= \int \lambda_c \frac{1}{B(\ba)} \prod_{l=1}^L \lambda_l^{\alpha_l -1 + \sum_{l'\in \mathcal{L}}\delta_{cl'}} d\bl \\
    &= \frac{B(\ba')}{B(\ba)} \int \text{Dir}(\ba') d\bl = \frac{B(\ba')}{B(\ba)},  \end{align*}
where $\alpha'_l=\alpha_l -1 + \sum_{l'\in \mathcal{L}} \delta_{cl'}$. Thus, the
posterior predictive distribution is also Categorically distributed as $c|\mathcal{L}\sim\text{Cat}(\bl')$ with $p_{c|\mathcal{L}}(c|\mathcal{L})= \frac{B(\ba')}{B(\ba)} = \lambda_c'$.  Only one label $l_s$ is added at each time $s$, thus using the recursive property of the Gamma function, $\Gamma(n+1) = n\Gamma(n)$, we see that 
\begin{align} \label{eq:ctoa} \notag
     \lambda_c' &= \frac{B(\ba')}{B(\ba)} = \frac{\Gamma( \alpha_0)}{\prod_{l=1}^L \Gamma(\alpha_l)}  \frac{\Gamma(\alpha_c + 1) \prod_{l=1\neq c}^{L}\Gamma(\alpha_l)}{\Gamma(\alpha_0 + 1)} \\
     &= \frac{\Gamma( \alpha_0)}{\prod_{l=1}^L \Gamma(\alpha_l)}  \frac{\alpha_c' \prod_{l=1}^{L} \Gamma(\alpha_l)}{\alpha_0' \Gamma(\alpha_0)} = \frac{\alpha_c'}{\alpha_0'}.
\end{align}

\begin{figure}[h]
    \centering
    \includegraphics[width=.48\textwidth]{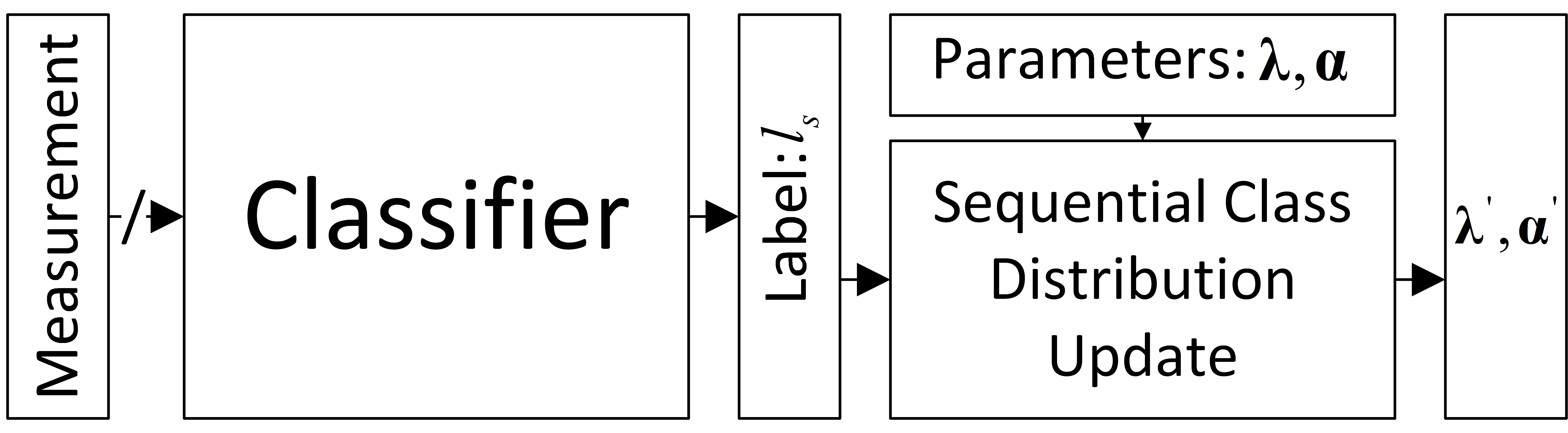}
    \caption{Block diagram of sequential classification state tracking.  A measurement is converted into a class label $l_s$ by means of a one-step classifier.  The class label is used in conjunction with the previous distribution parameters $\bl$ and $\ba$ to produce a new distribution parameter $\bl'$ and $\ba'$.}
    \label{fig:seq_class}
\end{figure}

Similar to occupancy grids, which capture the posterior marginal probabilities of occupancy for all grid cells, the posterior predictive class distribution generated by the sequential classification process is captured in classification maps (CMs).  We represent a CM as a set of distributions denoted as
\begin{align} \label{eq:cm_posterior_prob}
    \q = \{p_{c|\mathcal{L}}(c_r |\mathcal{L})\}_{r=1}^B,
\end{align}
where $c_r$ is the class state variable for the $r$th grid cell $g_r$.  Figure \ref{fig:seq_class} depicts the idea behind the proposed class state tracking process. The sequential class updating process takes each new class label $l_s$ as well as the previous distribution parameters $\bl$ and $\ba$ to produce the new distribution parameters $\bl'$ and $\ba'$, hence allowing us to successively update the parameter estimates for every new measurement.  

\section{Information-Theoretic Cost Functions}
\label{sec:info_metrics}
In this section, we develop a novel information-theoretic cost function that is used in evaluating the utility of sensing actions during navigation policy selection. This cost function, called the \textit{total information gain} at time $s$ $IG_T(s)$, is a convex combination of two individual information-theoretic cost functions, one for detection $IG_D(s)$ and one for classification $IG_C(s)$. We first present the detection cost function, derived with the sensor model and outputs from the occupancy grid estimation process.  The classification cost function is then presented, derived with the sequential classification process using the DCM. 

Letting $X$ and $Y$ be two random variables, and $H(\cdot)$ the Shannon entropy, then the information gain is defined as \cite{cover2012elements}
\begin{align} \label{eq:mutual_information} 
    I(X;Y) &= H(X) - H(X | Y),
\end{align}
which can be thought of as the reduction in uncertainty of the distribution on $X$ that the knowledge of random variable $Y$ would bring.  In the active perception problem, $H(X)$ can be thought of as the prior distribution on $X$, while $H(X|Y)$ can be viewed of as the posterior distribution on $X$ after observing random variable $Y$.  

In the context of our problem, both the detection and classification cost functions are closed-forms of the information gain, or mutual information, between a state variable $X$ (occupancy state and class state) and a random variable $Y$ that captures information about the latest measurements.  As such, they are both predictors of the information that is gained by taking a measurement from a grid cell given the current state of the grid cell.

In the following subsections, we will denote the set of grid cells observed at time $s$ by $\mathcal{G}$.

\subsection{Detection Cost Function}
We define the detection information gain at time $s$, $IG_D(s)$, as the sum of the mutual information between the occupancy state and the measurement random variable for all observed grid cells in $\mathcal{G}$.  This is stated mathematically as 
\begin{align} \label{eq:total_info_gain_det}
    IG_D(s) &\triangleq \sum_{g_i \in \mathcal{G}} I(b_i; j_{s,k}),    
\end{align}
where $b_i$ is the occupancy state variable for grid cell $g_i$ and $j_{s,k}$ is the random variable representing a measurement at time $s$ and range $k$.
The information gain for grid cell $g_{i}$ is given by,
\begin{align} \label{eq:info_gain_det} \notag
    I(b_{i};&j_{s,k}) = H(b_{i}) - H(b_{i} | j_{s,k}) \\ \notag
    &= -\mathbb{E}_{\mathcal{B}} \log p_{b}(b_{i}) - \mathbb{E}_{\mathcal{J}} \log p_{b|j}(b_{i} | j_{s,k}) \\
    &= -\sum_{b \in \mathcal{B}}p_{b}(b_{i})\log p_{b}(b_{i})  \\ \notag
    &\hspace{.2in} -\sum_{b \in \mathcal{B}}\sum_{j\in \mathcal{J}} p_{j|b}(j_{s,k}|b_{i})p_{b}(b_{i})\log \frac{p_{j|b}(j_{s,k}|b_{i})p_{b}(b_{i})}{p_{j}(j_{s,k})}.
\end{align}
where $\mathcal{J}=\{0,1\}$ is the set of possible values that realizations of $j_{s,k}$ can take, and $\mathcal{B}=\{0,1\}$ is the set of possible values that realizations of $b_i$ can take.

Using the occupancy grid estimation model in (\ref{eq:measurement_model}), the interaction between the occupancy state variable $b_i$ and the measurement random variable $j_{s,k}$ can be represented by,
\begin{align} \label{eq:detection_ig_meas_model} \notag
     p_{j|b}(j_{s,k}|b_i) &= {\left[(1-p_{\text{fa}})(1-b_i) +  (1-p_{\text{d}})b_i\right]}(1-j_{s,k}) \\
     &\hspace{.4in}+ {\left[p_{\text{fa}}(1-b_i) + p_{\text{d}}b_i\right]}j_{s,k},
\end{align}
where $p_{\text{d}}$ and $p_{\text{fa}}$ are the probabilities of detection and false alarm, respectively, for the physical detector that produces the measurement vectors $\j_s$.

We treat the output of the occupancy grid estimation $p_{b|\J}(b_i | \J_{s-1})$, i.e. the \textit{posterior estimate} from the previous time step, as the \textit{prior} $p_{b}(b_i)$ for time $s$ \cite{lindley2000stats}.  To simplify notation, we let $p_i=p_{b}(b_i=1)$. Now, using this together with (\ref{eq:detection_ig_meas_model}), and invoking the total probability, the marginal probability mass function for $j_{s,k}$ is,
\begin{align} \label{eq:marginal_j} \notag
     p_{j}(j_{s,k}) &= \sum_{\beta\in\mathcal{B}} p_{j|b}(j_{s,k} | b_i = \beta) p(b_i = \beta) \\ \notag
     &= p_{j|b}(j_{s,k} | b_i = 1)p_i + p_{j|b}(j_{s,k} | b_i = 0)(1-p_i) \\ \notag
     &= \left[(1-p_{\text{d}})(1-j_{s,k}) + p_{\text{d}}j_{s,k}\right]p_i \\ \notag
     & \hspace{.3in} + \left[(1-p_{\text{fa}})(1-j_{s,k}) + p_{\text{fa}}j_{s,k}\right](1-p_i) \\ \notag
     &= {\left[(1-p_{\text{fa}})(1-p_i) +  (1-p_{\text{d}})p_i\right]}(1-j_{s,k}) \\
     &\hspace{.4in} + {\left[p_{\text{fa}}(1-p_i) + p_{\text{d}}p_i\right]}j_{s,k},
\end{align}
Using this result, the prior and conditional entropy in (\ref{eq:info_gain_det}), become
\begin{align} \label{eq:entropy_prior}
    H(b_i) = -\left[ p_i\log p_i + (1-p_i)\log (1-p_i) \right].
\end{align}
and
\begin{align} \label{eq:entropy_cell} \notag
    &H(b_{i} | j_{s,k})  \\ \notag
    &=-\Big[ (1-p_{\text{fa}})\big(1-p_{i}\big)\log\frac{(1-p_{\text{fa}})\big(1-p_{i}\big)}{(1-p_{\text{d}})p_{i}+(1-p_{\text{fa}})\big(1-p_{i}\big)} \\ \notag
    &\hspace{.2in}  + p_{\text{fa}}\big(1-p_{i}\big)\log\frac{p_{\text{fa}}\big(1-p_{i}\big)}{p_{\text{d}} p_{i}+p_{\text{fa}}\big(1-p_{i}\big)}  \\ \notag
    &\hspace{.2in} + (1-p_{\text{d}})p_{i}\log\frac{(1-p_{\text{d}})p_{i}}{(1-p_{\text{d}})p_{i}+(1-p_{\text{fa}})\big(1-p_{i}\big)}  \\ \notag
    &\hspace{.2in} + p_{\text{d}}p_{i}\log\frac{p_{\text{d}}p_{i}}{p_{\text{d}} p_{i}+p_{\text{fa}}\big(1-p_{i}\big)} \Big] \\ \notag
    &= (1-p_{\text{fa}}) \big(1-p_{i}\big)\log\big[1+(1-p_{\text{d}})p_{i} \big] \\ \notag
    &\hspace{.4in} + p_{\text{fa}} \big(1-p_{i}\big)\log\big[1+p_{\text{d}} p_{i} \big] \\ \notag
    &\hspace{.2in} + (1-p_{\text{d}}) p_{i}\log\big[1+(1-p_{\text{fa}})\big(1-p_{i}\big) \big] \\
    &\hspace{.4in} + p_{\text{d}} p_{i}\log\big[1+p_{\text{fa}}\big(1-p_{i}\big)\big],
\end{align}
respectively. 

\noindent
Plugging (\ref{eq:detection_ig_meas_model}), (\ref{eq:marginal_j}), (\ref{eq:entropy_cell}), and (\ref{eq:entropy_prior}) into (\ref{eq:info_gain_det}) gives a closed-form expression for the detection information gain as
\begin{align} \label{eq:ig_cell} \notag
    I(b_{i};j_{s,k}) &= p_{i}\Big[ (1-p_{\text{d}})\log\big[1+(1-p_{\text{fa}})\big(1-p_{i}\big)\big]  \\ \notag
    &\hspace{.4in} + p_{\text{d}}  \log\big[1+p_{\text{fa}}\big(1-p_{i}\big)\big]  - \log p_{i}  \Big] \\ \notag
    &\hspace{.2in} + \big(1-p_{i}\big)\Big[ (1-p_{\text{fa}})\log\big[1+(1-p_{\text{d}})p_{i}\big]  \\
    &\hspace{.4in} + p_{\text{fa}} \log\big[1+p_{\text{d}} p_{i}\big]  - \log \big[1-p_{i}\big]  \Big].
\end{align}

\subsection{Classification Cost Function}
We define the classification information gain at time $s$ $IG_C(s)$ as the sum of the mutual information between the class state variable, $c_i$, and the Dirichlet distributed parameter vector, $\bl$, for all observed grids $g_i \in \mathcal{G}$. This is stated mathematically as 
\begin{align} \label{eq:total_info_gain_class}
    IG_C(s) &\triangleq \sum_{g_i \in \mathcal{G}} I(\bl;c_i).    
\end{align}
The distribution parameter vector $\bl$ for the Categorical distribution on $c_i$ in essence captures information about the latest measurements.

For a \textit{mixed-pair} of discrete scalar random variable $X$ and continuous random vector $\Y$, assuming they satisfy the sufficient conditions to be a \textit{good mixed-pair} \cite{nair2006entropy}, their mutual information is,
\begin{align} \label{eq:mixedmi} \notag
    I(\Y;X) &= h(\Y) - h(\Y|X) \\ \notag
    &= -\int p_{\y}(\y) \log{p_{\y}(\y)} d\y \\
    &\hspace{.3in} +  \sum_{x\in\mathcal{X}} \int p_{\y,x}(\y,x)\log{p_{\y|x}(\y|x)}d\y,
\end{align}
where $h(\cdot)$ is the differential entropy \cite{cover2012elements}.

Applying (\ref{eq:mixedmi}) to (\ref{eq:total_info_gain_class}), the mutual information $I(\bl;c)$ can be evaluated as 
\begin{align} \label{eq:info_gain_class} 
    I(\bl;c_i) &= h(\bl) - h(\bl | c_i) \\ \notag
    &= h(\bl) + \sum_{c_i=1}^L \int_{\Delta^L} p_{\bl,c_i}(\bl,c_i)\log{p_{\bl|c_i}(\bl|c_i)}d\bl.
\end{align}
The entropy of a Dirichlet distributed random vector is well-known \cite{ebrahimi2011information} and can be written as
\begin{align} \label{eq:direntropy}
    h(\bl) &= \log{B(\ba)} + (\alpha_0 - L)\psi(\alpha_0) - \sum_{l=1}^L (\alpha_l - 1)\psi(\alpha_l),
\end{align}
where $\psi(x)=\frac{d}{dx}\log{\Gamma(x)} = \frac{\Gamma'(x)}{\Gamma(x)}$ is the digamma function. 

To evaluate the conditional entropy term $h(\bl|c)$ in (\ref{eq:info_gain_class}), we use the fact that the Dirichlet distribution is the conjugate prior of the Categorical distribution. Thus, we can write
\begin{align} \label{eq:condentropy} \notag
    &h(\bl|c) = -\sum_{c=1}^L \int_{\Delta^L} p_{\bl,c}(\bl,c)\log{p_{\bl|c}(\bl|c)}d\bl \\ \notag
    &= -\sum_{c=1}^L \int_{\Delta^L} p_{c|\bl}(c|\bl)p_{\bl}(\bl)\log{p_{\bl|c}(\bl|c)}d\bl \\ \notag
    &= -\sum_{c=1}^L \frac{B(\ba')}{B(\ba)} \int_{\Delta^L}  \frac{1}{B(\ba')}  \prod_{l=1}^L \lambda_l^{\alpha' - 1} \log{\frac{1}{B(\ba')}\prod_{l=1}^L \lambda_l^{\alpha' - 1}} d\bl \\
    &= -\sum_{c=1}^L \frac{\alpha_c'}{\alpha_0'} E_{\bl'}\left[ \log{\frac{1}{B(\ba')}\prod_{l=1}^L \lambda_l^{\alpha' - 1}}  \right] = \sum_{c=1}^L \frac{\alpha_c'}{\alpha_0'} h(\bl') \\ \notag
    &= \sum_{c=1}^L \frac{\alpha_c'}{\alpha_0'} \left[ \log{B(\ba')} + (\alpha_0' - L)\psi(\alpha_0')  - \sum_{l=1}^L (\alpha_l' - 1)\psi(\alpha_l')\right]
\end{align}

Combining (\ref{eq:total_info_gain_class}), (\ref{eq:direntropy}), and (\ref{eq:condentropy}) provides the information gain for classification as
\begin{align} \label{eq:final_total_info_gain_class} 
    &IG_C(s) = \sum_{g_i \in \mathcal{G}} I(\bl;c_i) \\ \notag
    &= \sum_{g_i \in \mathcal{G}} \Big[ \log{B(\ba)} + (\alpha_0 - L)\psi(\alpha_0)    - \sum_{l=1}^L (\alpha_l - 1)\psi(\alpha_l) \\ \notag
    & - \sum_{c_i=1}^L \frac{\alpha_c'}{\alpha_0'} \Big( \log{B(\ba')}    + (\alpha_0' - L)\psi(\alpha_0') - \sum_{l=1}^L (\alpha_l' - 1)\psi(\alpha_l')\Big) \Big]. 
\end{align}

\subsection{Total Information Gain}
As previously stated, the total information gain is defined as the convex combination of the detection and classification information metrics.  To ensure that one information metric does not dominate the other at all times $s$ due to scaling, we normalize them by their respective maximal values, 
\begin{align}
    IG_T(s) = w_D\frac{IG_D(s)}{IG_{D_{\text{max}}}}  + w_C\frac{IG_C(s)}{IG_{C_{\text{max}}}} , && w_D + w_C = 1.
\end{align}
Note that for mutual information, the maximum value is attained when the prior distribution is uniform and the posterior distribution is fully deterministic (i.e., a delta function).  This convex weighting allows for different strategies to be employed by the system.  For example, at the beginning of a sortie, there will likely be insufficient information to consider classification for choosing sensing actions. In this case, we can assign a higher weight (e.g., $w_D=1$) for the detection while choosing a lower weight (e.g., $w_C = 1 - w_D = 0$) for the classification. In contrast, once most of the grid cells are observed, there may be little to no information left in performing target detection and localization, in which case we can choose the sensing actions solely  based on the classification  criterion by choosing $w_C=1$ and $w_D = 1 - w_C = 0$. 

\section{Trajectory Planning}
\label{sec:nav_policies}
In this section, we discuss three types of trajectory-planning policies for performing interactive sensing and navigation.  Recall that the action at each time step is the selection of the next location for the sensor platform.  For all policies, let $\a$ denote an action, and $\mathcal{A}_s$ be the set of all feasible actions the sensor can take under dynamical constraints at time step $s$.
The three policies choose sensing locations according to: a pre-determined lawn mower path; and maximizing a cost function for choosing $\a\in\mathcal{A}_s$ in a
greedy, and non-greedy manner.
\begin{figure}
    \centering
	\subfigure[Truth OG]{\includegraphics[width=0.23\textwidth]{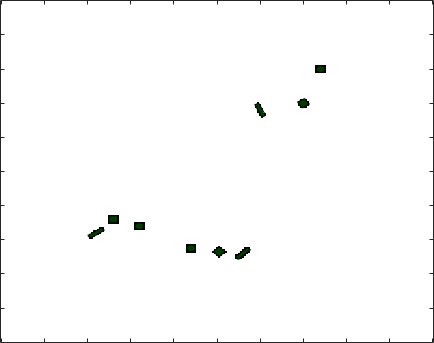} \label{fig:og_true}} %
	\subfigure[Truth CM]{\includegraphics[width=0.23\textwidth]{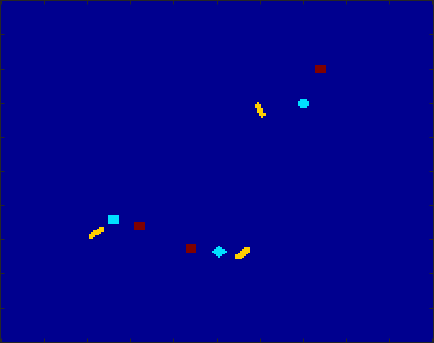} \label{fig:cm_true}}\\
	\subfigure[Lawn Mower OG]{\includegraphics[width=0.23\textwidth]{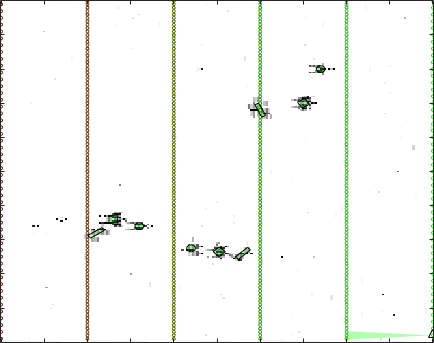} \label{fig:og_lawn}} %
	\subfigure[Lawn Mower CM]{\includegraphics[width=0.23\textwidth]{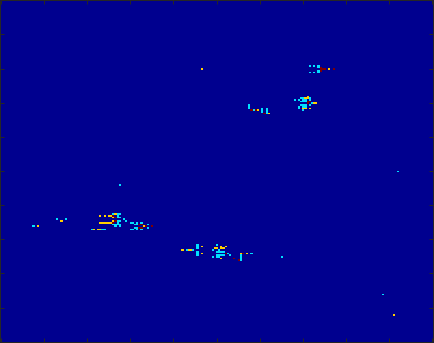} \label{fig:cm_lawn}}\\
	\subfigure[GPR-5 OG]{\includegraphics[width=0.23\textwidth]{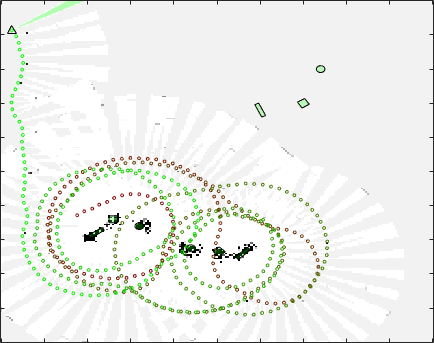} \label{fig:og_gpr}} %
	\subfigure[GPR-5 CM]{\includegraphics[width=0.23\textwidth]{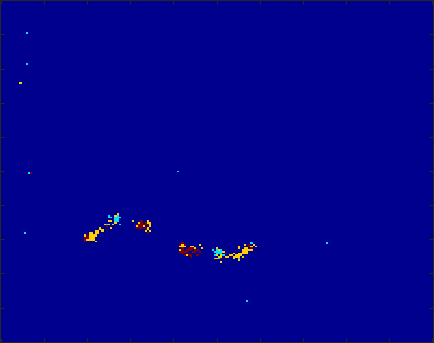} \label{fig:cm_gpr}}\\
	\subfigure[OG-DCM-5 OG]{\includegraphics[width=0.23\textwidth]{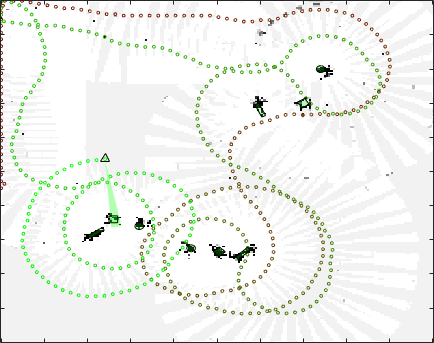} \label{fig:og_ours}} %
	\subfigure[OG-DCM-5 CM]{\includegraphics[width=0.23\textwidth]{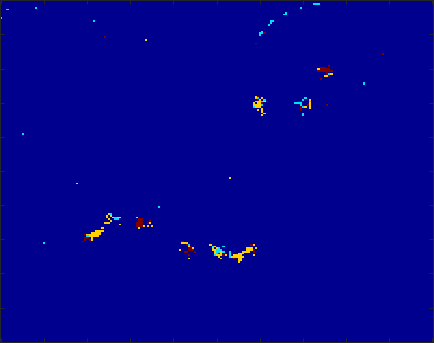} \label{fig:cm_ours}}
    \caption{Occupancy grids (OG) on left and classification maps (CM) on right. OG and CM shown for: the underlying truth, lawn mower, GPR-5, and OG-DCM-5. Cylinders are colored yellow, cubes light blue, and spheres red.  The deep blue color indicates no target. Each figure shows the same $50\times 50$ meter area.}
    \label{fig:all_maps}
\end{figure}

In many traditional underwater target detection and classification operations, a predetermined lawn mower path, as shown in Figure \ref{fig:og_lawn}, is used.  The sequence of prespecified actions $\a$ guarantees that each region in the search area is observed though at the  cost of potentially poor detection and classification accuracy due to inadequate viewing angles.

The greedy policy for action selection chooses the action as the one that maximizes the one-step \textit{reward}, or one-step navigation cost function, i.e.
\begin{align} \label{eq:greedy}
    \a^{\ast}_{s+1} = {\argmax}_{\a\in\mathcal{A}_s} R_s(\b_s, \c_s, \a),
\end{align}
where $R_s(\b_s, \c_s, \a) = IG_T(s)$ is the immediate reward for performing action $\a$ with the current state variable distributions at time $s$, where $\b_s$ and $\c_s$ are the occupancy state and the class state of the system at time $s$, respectively. 

The non-greedy policy for action selection chooses the next sensing action that maximizes the one-step reward, while also considering the reward from future actions along a finite horizon of $T$ future actions.  This policy, in essence, chooses the next sensing action that it believes will generate the largest cumulative reward given the current information available.  This type of problem is typically cast as a partially observable Markov decision problem (POMDP), which admits a number of solution methods \cite{bertsekas1999rollout, chong2009partially, goodson2017rollout}.  In this paper, the \textit{rollout policy} method \cite{chong2009partially} is used to solve the problem given our choice of a heuristic navigation cost function $IG_T(s)$.

Mathematically speaking, this policy can be described as choosing the sensing action that maximizes the one-step reward plus the \textit{expected reward-to-go} associated with a prespecified policy, typically a heuristic rule.  The decision rule for action selection used by the rollout policy is defined as
\begin{align}\label{eq:nongreedy}
    \a^{\ast}_{s+1} = {\argmax}_{\a\in\mathcal{A}_s}  R_s(\b_s, \c_s, \a) +  E_{s+1},
\end{align}
where $R_s(\b_s, \c_s, \a)$, $\b_s$, and $\c_s$ are the same as the greedy case, $E_{s+1}=\mathbb{E}\left[\sum_{i=s+1}^T R_i(\b_i, \c_i, \a_i )| \b_s, \c_s\right]$ is the expected reward-to-go, $R_i(\b_i, \c_i, \a_i) = IG_T(i)$ is the reward for performing action $\a_i$ with the current state variable distributions at time $i$, where $\b_i$ and $\c_i$ are the occupancy state and the class state of the system at time $i$, respectively, and $T$ is the length of the finite horizon.  

\section{Experimental Results}
\label{sec:simulations}

In this section, we present the simulation results from autonomous navigation experiments utilizing the action selection policies described in Section \ref{sec:nav_policies}. The experiments conducted in this section expose each action selection policy's ability to simultaneously perform detection, localization, and classification of targets, while exploring new areas.  For the remainder of this section, we call our proposed method for calculating $IG_T(s)$ the occupancy grid Dirichlet Categorical model (OG-DCM).

For each of the policies that use a navigation cost function (greedy and non-greedy policies), we benchmark OG-DCM to that of a Gaussian process regression (GPR) \cite{bai2018inference, hollinger2013active} for estimating the information gain expected from both the detection and classification components. We trained the Gaussian process with a Mat{\'e}rn kernel function \cite{bai2018inference}, augmented input vectors, and K-D trees \cite{hollinger2013active} to find the nearest $100$ neighbors for forming the covariance matrix on a per-cell basis. The augmented input vectors comprise the occupancy and classification state of the grid cell. The ground truth outputs used to train the Gaussian process were the sum of the detection and classification information gains, calculated as the difference between the entropy of prior and posterior state distributions for detection and classification, respectively \cite{cover2012elements}.  The total information gain is approximated by the Gaussian process regression following the Gaussian process upper bound confidence algorithm \cite{bai2018inference,srinivas2010gaussian}, and is given by
\begin{align*}
    IG_{GPR}(s) &= \mu(x) + \beta \sigma(x),
\end{align*}
where $\beta$ is the tradeoff parameter between exploration and exploitation, $\mu(x)$ and $\sigma(x)$ are the predicted mean and variance, respectively, derived from Gaussian process regression.

Experiments were run for the three different action selection policies under different configurations (different cost functions for the greedy and non-greedy policies), totaling 7 different experiments.  One accounts for the lawn mower policy, and six others account for different finite horizon lengths of $T=0$ (greedy), $T=5$ and $T=10$ for both GPR and OG-DCM.  We denote the finite horizon policies as GPR-$T$ and OG-DCM-$T$, i.e., GPR-0 and OG-DCM-0 are for $T=0$, etc.



\subsection{Experimental Data \& Description}
In our active perception problem,  a sonar platform is used to mimic the behavior of an autonomous underwater vehicle (AUV) that is searching littoral zones for mine-like underwater targets.  The system is equipped with multiple (11) hydrophones arranged in a uniform linear array (ULA), all pointing slightly downwards from horizontal (positive depression angle). The transmitted waveform was a linearly frequency modulated (LFM) chirp with center frequency $f_c=80$ kHz, bandwidth $BW=20$ kHz, and sampling frequency $f_s=60$ kHz.  The ULA has a $7^{\circ}$  beamwidth with an interrogation range up to tens of meters.  The sensor is attached to a platform, which is 10 meters above the seafloor.  

The experiments use simulated side-looking sonar (SLS) that directs acoustic radiation to the starboard side of the AUV. The sonar data is generated by the Personal Computer Shallow Water Acoustic Toolset (PC SWAT) simulation tool developed at the Naval Surface Warfare Center Panama City (NSWC PCD).  PC SWAT is the cutting-edge, physics-based sonar simulator that models  scattering from the target by a combination of the Kirchhoff approximation and the geometric theory of diffraction. Propagation of sound into a marine sediment with ripples is described by an application of Snell's law and second order perturbation theory in terms of Bragg scattering \cite{sammelmann2001propagation}.  PC SWAT has been used to produce simulations providing \textit{exemplar} template measurements that match real data generated by real shallow water sonar systems \cite{underwater2015serdp}.

The stave data generated by PC SWAT was fed through an adaptive coherence estimator (ACE) detector \cite{scharf1996adaptive, kraut2001adaptive, kraut2005adaptive} to produce a single beamformed measurement vector.  This beamformed vector is thresholded at a predetermined value to provide the measurement vector $\j_s$. The threshold is chosen to yield a desired $p_{\text{fa}}$ and hence $p_{\text{d}}$.

A total of 500 pings (actions) with $1$ meter ping separation were simulated for each experiment. Nine targets, in three clusters of three targets, are proud on the seafloor within the $50\times 50$ meter search field. Medium sandy bottom (clutter) was used in all the experiments.   Each cluster contains a cylindrical target that is $2$ meters long with a radius of $0.25$ meters, a cube target of $1$ meter in each dimension, and a partially hollow sphere with $1$ meter radius. Thus, the total number of classes is $L=3$.  The spatial orientation of the three clusters can be seen in the classification map (CM) of Figure \ref{fig:cm_true}, where the cylinders are color-coded yellow, the cubes light blue, and the spheres red.  The deep blue color indicates no target.

The one-step classifier used in these experiments is the modified matched subspace classifier (MMSC) \cite{hall2018underwater}, as it has been shown to be very successful in classifying underwater objects.  The MMSC uses sparse coding and dictionary learning, and has many desirable properties including the ability to use any learned subspace dictionaries, and incremental updating of the dictionary matrices when operating in new measurements.

At each time step, a sensing action in the form of selecting and moving to the next position to collect a measurement is taken.  The measurement is collected and used to update the occupancy grid and classification map. The next location from which to take a measurement was chosen according to (\ref{eq:greedy}) and (\ref{eq:nongreedy}) for the greedy and non-greedy policies, respectively.  Each reward, $R(\b_s, \c_s, \a) = IG_T(s)$ for OG-DCM and $R(\b_s, \c_s, \a) = IG_{GPR}(s)$ for GPR, was calculated by first generating the ping from the new location with PC SWAT, then performing the occupancy grid estimation and classification map estimation, and finally calculating the reward for that action.   The non-greedy policy was evaluated to time step $s+T$ for $T\in\{0,5,10\}$, i.e., each decision involves considering $T$ time steps into the future.

\subsection{Evaluation Metrics}
To evaluate the performance of each policy, three different metrics are used, as detailed below.  For all metrics, with the exception of the percentage of grid cells observed, we evaluate the performance for detection (occupancy grids) and classification (classification maps) separately to better illustrate the strengths and weaknesses of each.  In the following, $\t$ represents the true set of distributions, either true occupancy $\b$ or true classification $\c$, and $\e$ represents the estimated set of distributions, either occupancy grid $\p$ or classification map $\q$. Only like pairs are compared, i.e., $\b$ and $\p$ or $\c$ and $\q$.  The true distributions are formed from delta functions (e.g., $p_{b|\J}(b_r|\J_S) = [1, 0]$ if a cell is occupied, and $p_{c|\mathcal{L}}(c_r|\mathcal{L}) = [0, 0, 1, 0]$ if a cell is of class 3).  
\begin{enumerate}
    \item Similarity between the true distribution $\t$ to that of the estimated distribution $\e$:
    \begin{align*}
        \rho &= \frac{\langle \t,\e\rangle_{F}}{||\t||_{F}^{2}||\e||_{F}^{2}},
    \end{align*}
    where $||\cdot||_F$ is the Frobenius norm, and $\langle \cdot, \cdot \rangle_F$ is the Frobenius inner product.  For calculating this metric, we form $\t$ and $\e$ into matrices by making each row the vectorized form of the distribution for each grid cell. Clearly, $0 \leq \rho \leq 1$, and $\rho=1$ when $\t = \e$
    
    \item Sum of the Jensen-Shannon distance (SJSD)  $D_{\text{JS}}(t_i||e_i)$ \cite{lin1991divergence} over all $i=1,\hdots,B$ grid cells:
    \begin{align*}
        &\text{SJSD} = \sum_{r} D_{\text{JS}}(t_i||e_i)  \\
        &= \sum_{i} \frac{1}{2}D_{\text{KL}}(t_i||m_i) + \frac{1}{2} D_{\text{KL}}(e_i||m_i)\\
        &= -\frac{1}{2}\sum_{i} \Big[ \sum_{x\in\mathcal{X}} t_i(x) \log \Big( \frac{m_i(x)}{t_i(x)} \Big) + e_i(x) \log \Big( \frac{m_i(x)}{e_i(x)} \Big)\Big],
    \end{align*}
    where $m_i(x)=\frac{1}{2}\big(t_i(x)+e_i(x)\big)$, $t_i(x)$ and $e_i(x)$ are the distributions for grid cell $g_i$ evaluated at point $x$, and $D_{\text{KL}}(\cdot, \cdot)$ is the Kullback-Leibler (KL) divergence \cite{lin1991divergence}.  The Jensen-Shannon distance is used in favor of the KL divergence as it is symmetric, positive, and always finite. The maximum value of SJSD is $\log(2)\times B$, with smaller values indicating that $\t$ and $\e$ are similar and $\t=\e$ when SJSD is 0.
    \item Percentage of the grid cells in the map that are observed during the experiment. This measure provides the efficiency with which the AUV is able to explore new areas while estimating the occupancy grid. 
\end{enumerate}

The true distributions are illustrated in Figure \ref{fig:og_true} and \ref{fig:cm_true}.  In Figure \ref{fig:og_true}, occupied grid cells are black while empty grid cells are white.  In Figure \ref{fig:cm_true}, each grid cell is color-coded according to the non-zero component of its distribution (i.e., cylindrical class grid cells are yellow, etc.). 

\subsection{Results and Discussion}
For each type of non-deterministic policy, i.e., greedy, and non-greedy, $20$ different trials were conducted where the starting locations and headings of the AUV were randomly chosen in each trial.  The lawn mower policy was only executed once.  Table \ref{tab:results} gives the mean values of SJSD, $\rho$, and the percentage of grid cells observed after 500 sensing actions. Bold values in each column of the table represent the best performance for the metric associated with that column.
\begin{table}
\centering
\caption{SJSD and $\rho$ for detection (Det.) and classification (Class.), and \% of grid cells seen for different navigation policies after $500$ sensing actions. \textbf{Bold} values indicate best performance per metric.}
\resizebox{\columnwidth}{!}{
\begin{tabular}{c|c|c|c|c|c} \toprule
 Policy & \% Seen & SJSD Det. & SJSD Class. & $\rho$ Det. & $\rho$ Class.  \\
     \midrule
 Lawn Mower  & \textbf{0.97} & \textbf{157.5}    & 69.2          & \textbf{0.52} & 0.58         \\
 \hline
 GPR-0       & 0.10          & 764               & 66.5          & 0.22          & 0.61         \\
 GPR-5       & 0.45          & 542               & 68.6          & 0.26          & 0.58         \\
 GPR-10      & 0.41          & 558.1             & 70.83         & 0.15          & 0.57         \\
 \hline
 OG-DCM-0    & 0.35          & 604.7             & 62.7          & 0.38          & 0.64         \\
 OG-DCM-5    & 0.62          & 442.1             & 57.7          & 0.47          & 0.67         \\
 OG-DCM-10   & 0.62          & 427.1             & \textbf{48.7} & \textbf{0.52} & \textbf{0.75}\\
\end{tabular}
}
\label{tab:results}
\end{table}

As seen from these results the OG-DCM non-greedy policy with a 10 step finite horizon outperformed all other non-deterministic policies in all metrics.  The lawn mower policy outperformed the OG-DCM non-greedy 10 step policy in the number of observed grid cells and the SJSD for detection.  These two metrics can be thought of as measuring the same information, as an increase in the number of grid cells seen implies a relatively good estimate of occupancy for any reliable detector.
The policies that use OG-DCM generally outperformed the policies that use GPR.  In fact, there are only a few cases where even the greedy policy using OG-DCM performed worse than the best non-greedy policy using GPR.  This shows that the proposed OG-DCM method provides better estimation of the information gain for future sensing actions than that of GPR.

The OGs and CMs illustrated in Figures \ref{fig:og_lawn}-\ref{fig:cm_ours} show one realization of the results for the lawn mower, and non-greedy policy experiments.  Realizations for finite horizons of $T=0$ and $T=10$ have been omitted due to space limitation.  The policies that use GPR wound up falling into local minima, i.e., over-observing a particular region without extracting any new detection or classification information, and thus reduced the overall efficiency of the system. Again, as can be observed the OG-DCM method generally provided much better classification results when compared with the GPR results. 

\begin{figure}
    \centering
	\subfigure[\% of grid cells seen]{\includegraphics[width=0.48\textwidth]{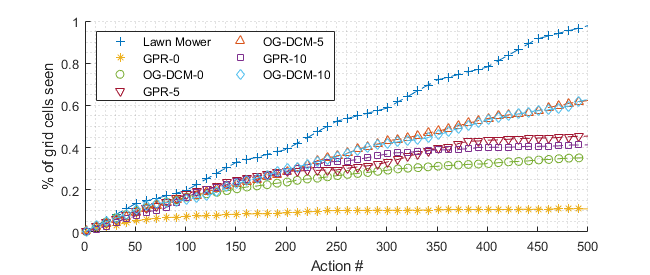} \label{fig:percent_seen}} \\
	\subfigure[SJDS for detection]{\includegraphics[width=0.48\textwidth]{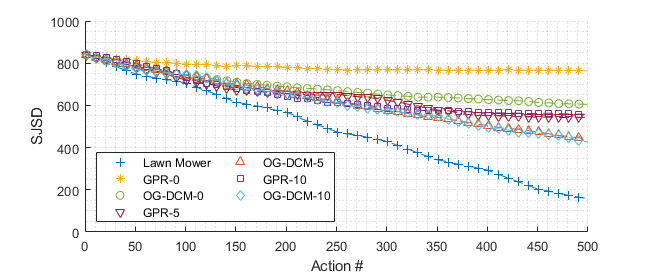} \label{fig:sjds_det}}\\
	\subfigure[SJDS for classification]{\includegraphics[width=0.48\textwidth]{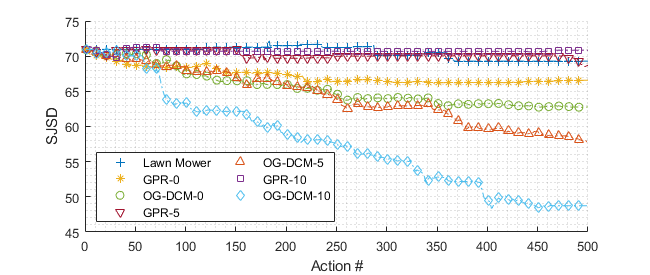} \label{fig:sjds_class}}\\
	\subfigure[$\rho$ for detection]{\includegraphics[width=0.48\textwidth]{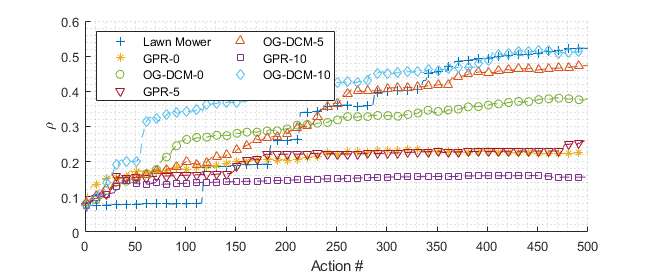} \label{fig:rho_det}} \\
	\subfigure[$\rho$ for classification]{\includegraphics[width=0.48\textwidth]{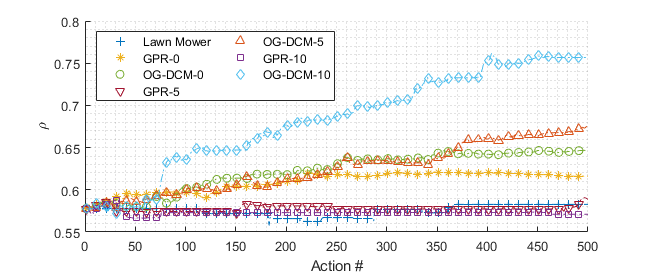} \label{fig:rho_class}}
    \caption{Performance plots for each of the navigation policies used in experiments.}
    \label{fig:performance_plots}
\end{figure}

Finally, to compare the temporal evolution of these metrics during the navigation the mean value of each metric is plotted in Figures \ref{fig:sjds_det}-\ref{fig:percent_seen} against the action number for each case.  A major take away, illustrated in Figure \ref{fig:sjds_class}, \ref{fig:rho_det}, and \ref{fig:rho_class}, is that OG-DCM-10 not only outperforms the other policies, but does so in a relatively small number of sensing actions.  From these results, one can conclude that as more sensing actions are taken, and more grid cells are observed, the OG-DCM-10 outperforms all other policies in all metrics.

\section{Conclusion} \label{sec:conc}
An autonomous navigation system using a novel information-theoretic cost function is proposed based on the outputs of two state tracking algorithms, for object detection and object classification.  This navigation cost function provides a way to compare multiple locations from which a sensor can take measurements and declare which of those locations provide maximal information gain for updating state variable estimates.  The performance of three navigation policies, using the proposed cost function, were evaluated and compared to the ground truth.  The experimental results show that the use of the proposed information-theoretic cost function along with the non-greedy policy produces the most accurate target classification and occupancy estimates while observing more of the map in the same duration of time.


\bibliographystyle{IEEEtran}
\bibliography{refs}

\begin{thebibliography}{10}
\providecommand{\url}[1]{#1}
\csname url@samestyle\endcsname
\providecommand{\newblock}{\relax}
\providecommand{\bibinfo}[2]{#2}
\providecommand{\BIBentrySTDinterwordspacing}{\spaceskip=0pt\relax}
\providecommand{\BIBentryALTinterwordstretchfactor}{4}
\providecommand{\BIBentryALTinterwordspacing}{\spaceskip=\fontdimen2\font plus
\BIBentryALTinterwordstretchfactor\fontdimen3\font minus
  \fontdimen4\font\relax}
\providecommand{\BIBforeignlanguage}[2]{{%
\expandafter\ifx\csname l@#1\endcsname\relax
\typeout{** WARNING: IEEEtran.bst: No hyphenation pattern has been}%
\typeout{** loaded for the language `#1'. Using the pattern for}%
\typeout{** the default language instead.}%
\else
\language=\csname l@#1\endcsname
\fi
#2}}
\providecommand{\BIBdecl}{\relax}
\BIBdecl

\bibitem{bajcsy1988active}
R.~Bajcsy, ``Active perception,'' \emph{Proceedings of the IEEE}, vol.~76,
  no.~8, pp. 966--1005, 1988.

\bibitem{whaite1997autonomous}
P.~Whaite and F.~P. Ferrie, ``Autonomous exploration: Driven by uncertainty,''
  \emph{IEEE Trans. Pattern Analysis and Machine Intelligence}, vol.~19, no.~3,
  pp. 193--205, 1997.

\bibitem{hollinger2013active}
G.~A. Hollinger, B.~Englot, F.~S. Hover, U.~Mitra, and G.~S. Sukhatme, ``Active
  planning for underwater inspection and the benefit of adaptivity,'' \emph{The
  International Journal of Robotics Research}, vol.~32, no.~1, pp. 3--18, 2013.

\bibitem{bourgault2002information}
F.~Bourgault, A.~A. Makarenko, S.~B. Williams, B.~Grocholsky, and H.~F.
  Durrant-Whyte, ``Information based adaptive robotic exploration,'' in
  \emph{IEEE/RSJ Int. Conf. Intelligent Robots and Systems}, vol.~1.\hskip 1em
  plus 0.5em minus 0.4em\relax IEEE, 2002, pp. 540--545.

\bibitem{bai2018inference}
S.~Bai, J.~Wang, K.~Doherty, and B.~Englot, ``Inference-enabled
  information-theoretic exploration of continuous action spaces,'' in
  \emph{Robotics Research}.\hskip 1em plus 0.5em minus 0.4em\relax Springer,
  2018, pp. 419--433.

\bibitem{julian2014mutual}
B.~J. Julian, S.~Karaman, and D.~Rus, ``On mutual information-based control of
  range sensing robots for mapping applications,'' \emph{Int. J. Robotics
  Research}, vol.~33, no.~10, pp. 1375--1392, 2014.

\bibitem{elfes1989using}
A.~Elfes, ``Using occupancy grids for mobile robot perception and navigation,''
  \emph{Computer}, vol.~22, no.~6, pp. 46--57, 1989.

\bibitem{thrun2003learning}
S.~Thrun, ``Learning occupancy grid maps with forward sensor models,''
  \emph{Autonomous Robots}, vol.~15, no.~2, pp. 111--127, 2003.

\bibitem{robbiano2020bayesian}
C.~Robbiano, E.~K.~P. Chong, L.~L. Scharf, M.~R. Azimi-Sadjadi, A.~Pezeshki,
  and S.~Ahmadinia, ``Bayesian learning of occupancy grids,'' 2019, preprint,
  \url{https://arxiv.org/abs/1911.07915}.

\bibitem{murphy2012machine}
K.~P. Murphy, \emph{Machine learning: a probabilistic perspective}.\hskip 1em
  plus 0.5em minus 0.4em\relax MIT press, 2012.

\bibitem{cover2012elements}
T.~M. Cover and J.~A. Thomas, \emph{Elements of Information Theory}.\hskip 1em
  plus 0.5em minus 0.4em\relax John Wiley \& Sons, 2012.

\bibitem{tu2014dirichlet}
S.~Tu, ``The dirichlet-multinomial and dirichlet-categorical models for
  bayesian inference,'' \emph{Computer Science Division, UC Berkeley}, 2014.

\bibitem{bishop2006pattern}
C.~M. Bishop, \emph{Pattern recognition and machine learning}.\hskip 1em plus
  0.5em minus 0.4em\relax springer, 2006.

\bibitem{goodfellow2016deep}
I.~Goodfellow, Y.~Bengio, and A.~Courville, \emph{Deep learning}.\hskip 1em
  plus 0.5em minus 0.4em\relax MIT press, 2016.

\bibitem{lindley2000stats}
\BIBentryALTinterwordspacing
D.~V. Lindley, ``The philosophy of statistics,'' \emph{Journal of the Royal
  Statistical Society. Series D (The Statistician)}, vol.~49, no.~3, pp.
  293--337, 2000. [Online]. Available:
  \url{http://www.jstor.org/stable/2681060}
\BIBentrySTDinterwordspacing

\bibitem{nair2006entropy}
C.~Nair, B.~Prabhakar, and D.~Shah, ``On entropy for mixtures of discrete and
  continuous variables,'' \emph{arXiv preprint cs/0607075}, 2006,
  \url{https://arxiv.org/abs/cs/0607075}.

\bibitem{ebrahimi2011information}
N.~Ebrahimi, E.~S. Soofi, and S.~Zhao, ``Information measures of dirichlet
  distribution with applications,'' \emph{Applied Stochastic Models in Business
  and Industry}, vol.~27, no.~2, pp. 131--150, 2011.

\bibitem{bertsekas1999rollout}
D.~P. Bertsekas and D.~A. Castanon, ``Rollout algorithms for stochastic
  scheduling problems,'' \emph{J. Heuristics}, vol.~5, no.~1, pp. 89--108,
  1999.

\bibitem{chong2009partially}
E.~K. Chong, C.~M. Kreucher, and A.~O. Hero, ``Partially observable markov
  decision process approximations for adaptive sensing,'' \emph{Discrete Event
  Dynamic Systems}, vol.~19, no.~3, pp. 377--422, 2009.

\bibitem{goodson2017rollout}
J.~C. Goodson, B.~W. Thomas, and J.~W. Ohlmann, ``A rollout algorithm framework
  for heuristic solutions to finite-horizon stochastic dynamic programs,''
  \emph{Eur. J. Operational Research}, vol. 258, no.~1, pp. 216--229, 2017.

\bibitem{srinivas2010gaussian}
N.~Srinivas, A.~Krause, S.~Kakade, and M.~Seeger, ``Gaussian process
  optimization in the bandit setting: no regret and experimental design,'' in
  \emph{Proceedings of the 27th International Conference on International
  Conference on Machine Learning}, 2010, pp. 1015--1022.

\bibitem{sammelmann2001propagation}
G.~S. Sammelmann, ``Propagation and scattering in very shallow water,'' in
  \emph{MTS/IEEE Oceans 2001. An Ocean Odyssey. Conference Proceedings (IEEE
  Cat. No. 01CH37295)}, vol.~1.\hskip 1em plus 0.5em minus 0.4em\relax IEEE,
  2001, pp. 337--344.

\bibitem{underwater2015serdp}
R.~Lim, ``Data and processing tools for sonar classification of underwater
  uxo,'' \emph{SERDP MR-2230}, 2015.

\bibitem{scharf1996adaptive}
L.~L. Scharf and L.~T. McWhorter, ``Adaptive matched subspace detectors and
  adaptive coherence estimators,'' in \emph{Conf. Rec. 13th Asilomar Conf.
  Signals, Systems and Computers}.\hskip 1em plus 0.5em minus 0.4em\relax IEEE,
  1996, pp. 1114--1117.

\bibitem{kraut2001adaptive}
S.~Kraut, L.~L. Scharf, and L.~T. McWhorter, ``Adaptive subspace detectors,''
  \emph{IEEE Trans. Signal Processing}, vol.~49, no.~1, pp. 1--16, 2001.

\bibitem{kraut2005adaptive}
S.~Kraut, L.~L. Scharf, and R.~W. Butler, ``The adaptive coherence estimator: A
  uniformly most-powerful-invariant adaptive detection statistic,'' \emph{IEEE
  Trans. Signal Processing}, vol.~53, no.~2, pp. 427--438, 2005.

\bibitem{hall2018underwater}
J.~J. Hall, M.~R. Azimi-Sadjadi, S.~G. Kargl, Y.~Zhao, and K.~L. Williams,
  ``Underwater unexploded ordnance (uxo) classification using a matched
  subspace classifier with adaptive dictionaries,'' \emph{IEEE Journal of
  Oceanic Engineering}, no.~99, pp. 1--14, 2018.

\bibitem{lin1991divergence}
J.~Lin, ``Divergence measures based on the {S}hannon entropy,'' \emph{IEEE
  Trans. Information theory}, vol.~37, no.~1, pp. 145--151, 1991.

\end{thebibliography}

\begin{IEEEbiography}[{\includegraphics[width=1in, height=1.25in, clip,keepaspectratio]{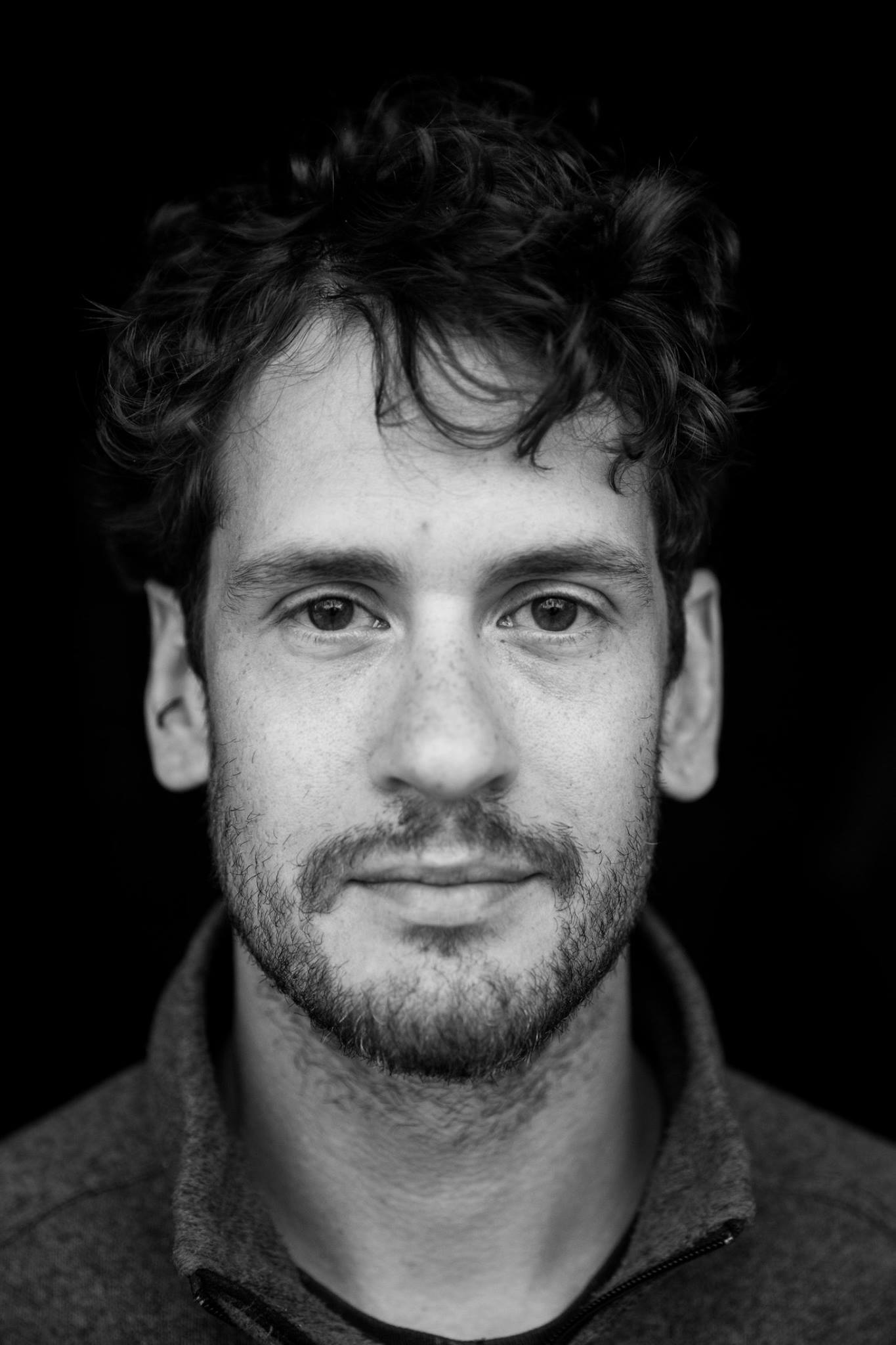}}]{Christopher~Robbiano} Christopher Robbiano received a BS in physics and a BS in electrical engineering, and a MS in electrical engineering from Colorado State University in 2011 and 2017, respectively.  He is currently a PhD candidate in electrical engineering at Colorado State University, where his research has been in the areas of detection and classification tasks in sonar applications.  He currently works as a research scientist at Information System Technologies Incorporated, and previously worked as a digital design engineer at Broadcom.  His research interests include machine learning, autonomous systems, statistical signal processing, and applications of the aforementioned topics to resource constrained embedded platforms.
\end{IEEEbiography}
\vskip -2\baselineskip plus -1fil
\begin{IEEEbiography}[{\includegraphics[width=1in, height=1.25in, clip,keepaspectratio]{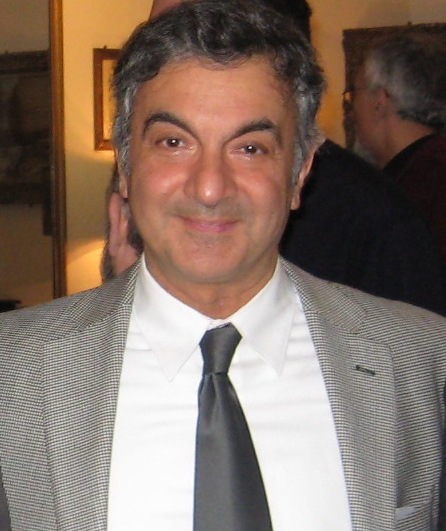}}]{Mahmood~R.~Azimi-Sadjadi}
Dr. Azimi-Sadjadi received his M.S. and Ph.D. degrees from the Imperial College of Science \& Technology, University of London, England in 1978 and 1982, respectively, both in Electrical Engineering with specialization in Digital Signal/Image Processing. 

He is currently a full professor at the Electrical and Computer Engineering Department at Colorado State University (CSU). He is also serving as the director of the Digital Signal/Image Laboratory at CSU.   His main areas of interest include statistical signal and image processing, machine learning and adaptive systems, target detection, classification and tracking, sensor array processing, and distributed sensor networks.  

Dr. Azimi-Sadjadi served as an Associate Editor of the IEEE Transactions on Signal Processing and the IEEE Transactions on Neural Networks.   He is a Life Member of the IEEE. 
\end{IEEEbiography}
\vskip -2\baselineskip plus -1fil
\begin{IEEEbiography}[{\includegraphics[width=1in, height=1.25in, clip,keepaspectratio]{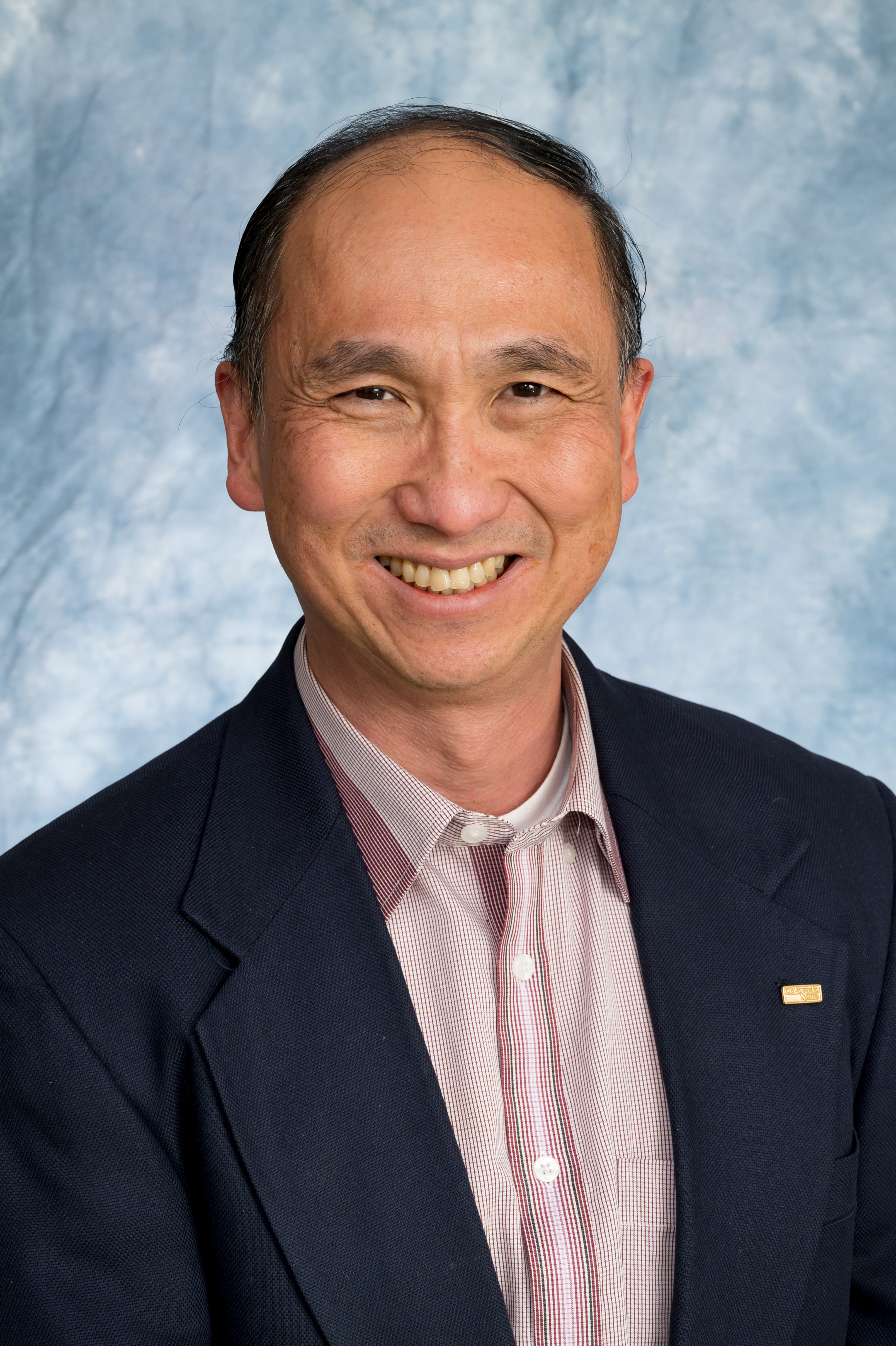}}]{Edwin~K.~P.~Chong}
Edwin K. P. Chong (F'04) received the B.E. degree with First Class Honors from the University of Adelaide, South Australia, in 1987; and the M.A. and Ph.D. degrees in 1989 and 1991, respectively, both from Princeton University, where he held an IBM Fellowship. He joined the School of Electrical and Computer Engineering at Purdue University in 1991, where he was named a University Faculty Scholar in 1999. Since August 2001, he has been a Professor of Electrical and Computer Engineering and Professor of Mathematics at Colorado State University. He coauthored the best-selling book, \emph{An Introduction to Optimization} (4th Edition, Wiley-Interscience, 2013). 

Prof. Chong received the NSF CAREER Award in 1995 and the ASEE Frederick Emmons Terman Award in 1998. He was a co-recipient of the 2004 Best Paper Award for a paper in the journal \emph{Computer Networks}. In 2010, he received the IEEE Control Systems Society Distinguished Member Award. He was the founding chairman of the IEEE Control Systems Society Technical Committee on Discrete Event Systems, and served as an IEEE Control Systems Society Distinguished Lecturer. He was a Senior Editor of the \textsc{IEEE Transactions on Automatic Control}.  He was the General Chair for the 2011 Joint 50th IEEE Conference on Decision and Control and European Control Conference.  He has served as a member of the IEEE Control Systems Society Board of Governors and as Vice President for Financial Activities until 2014. He currently serves as President (2017).
\end{IEEEbiography}

\end{document}